\documentstyle[12pt]{article}
\newcommand{\newc}{\newcommand}
\newc{\beq}    {\begin{equation}}
\newc{\eeq}    {\end{equation}}
\newc{\beqa}    {\begin{eqnarray}}
\newc{\eeqa}    {\end{eqnarray}}
\newc{\st}    {\stackrel}
\newc{\stg}    {\st{>}{\sim}}
\newc{\stl}    {\st{<}{\sim}}

\def\PLB{{\em Phys. Lett.}  {\bf B}}

\begin{document}

\begin{titlepage}
\title{ INFLATION AND INVERSE SYMMETRY BREAKING}
%\thanks{ e-mail address: leejw@chiak.kaist.ac.kr}
\author{ Jae-weon Lee and In-gyu Koh  \\ \it Department of Physics,
 \\ \it Korea Advanced Institute of Science and Technology,
\\   \it 373-1, Kusung-dong, Yusung-ku, Taejeon, Korea  \\
\\
}

\date{}
\setcounter{page}{1}
\maketitle
\vspace{-9cm}
%\hfill KAIST-CHEP-100/95
\vspace{10cm}

\hspace{.5cm}An inflation model with inverse symmetry
 breaking of two scalar fields is proposed.
Constraints on the parameters for
 a successful inflation are obtained.
In general the inequality
 $\lambda_1\ll g <\lambda_2$ should
be satisfied, where $\lambda_{1,2}$ and $g$ are
the coupling constants for self interaction and mutual
interaction of two scalar fields respectively.
An example with $SU(5)$ GUTs phase transition
and numerical study are presented.
This model introduce a new mechanism for the onset of inflation.
\maketitle

\vspace{3cm} PACS numbers: 98.80.Cq
\end{titlepage}

\newpage

\vskip 5mm
\section*{I.  INTRODUCTION}
\hspace{.5cm} Various inflation models\cite{guth} have been proposed
 to solve the horizon and the flatness problems of the standard Big
 Bang cosmology.  However, in relation to particle physics each
 model has its own problems to be solved.  Therefore, reconciling
the inflation models with particle physics
 is an important subject of the modern cosmology.

Since the upper bound on the
 inflation energy scale is about  the 
GUTs(Grand Unified Theories) scale\cite{liddle}, it is natural
 to search for the inflation during the GUTs phase transition. 
However, the original inflation model has graceful-exit
 problem\cite{guth} and `new' inflation model with the GUTs
 nonsinglet
 fields leads to too strong density fluctuation\cite{new}.  
As a solution to this problem the model with the GUTs singlet
 inflaton coupled with SU(5) Higgs was suggested\cite{su5}. 

Generally, the smallness of the coupling constants required
for the small density perturbation prevents
inflaton fields from obtaining thermal-equilibrium,
while chaotic inflation model\cite{chaotic} uses  
this non-equilibrium states to give the initial conditions
for the inflaton fields.

Though many aspects of the phase transition theory have already 
been used for the various inflation models, field theory has
 still other mechanisms of the  phase transition to be studied 
in the context of inflation. 

In this paper, an inflation model with `inverse
 symmetry breaking' is investigated.
Inverse symmetry breaking\cite{isb} is a phenomenon that the
 symmetry broken at a higher temperature is restored at a lower
 temperature, contrary to the ordinary
phase transitions.
The phenomenon has been applied to solving the
 monopole problem
 by allowing the temporary breaking of the U(1) gauge
 symmetry\cite{pi}.
Similar phenomenon called anti-restoration appears in
 some global SUSY theories\cite{anti}.

Our model is a kind of two fields inflation 
models\cite{two} where generally  an additional scalar 
field besides inflaton  is introduced to complete
 the inflation and/or give an appropriate density perturbations. 
For example, in the `hybrid' or `false vacuum'
 inflation model\cite{hybrid} the additional scalar
 field gives the inflaton extra masses which
 make the inflaton roll down and end the inflation.

The inflaton potential in our model is similar to that 
in the hybrid model, but the detailed features of the
 phase transition are very different.
In our model the phase transition of a scalar field 
(say $\phi_2$, for example GUTs Higgs) is responsible for the beginning
 of the inflation driven by a gauge singlet
 inflaton(say $\phi_1$) rather
 than the ending of the inflation. Moreover, the additional
 field($\phi_2$) 
is in the true vacuum rather than the false vacuum during 
the inflation.    

In sec. II, we review inverse symmetry breaking and derive
the conditions for the phenomenon.
In sec. III,  the constraints for the successful inflation
is derived.
In sec. IV,  an application with  $SU(5)$ GUTs model
and numerical study are presented.
Sec. V  contains discussions.

\vskip 2.5mm
\bigskip

\section*{II.  INVERSE SYMMETRY BREAKING }

\hspace{.5cm}
In this section we review inverse symmetry breaking
and conditions required for it.
 Consider a following potential which
 is a simple example of inverse symmetry breaking.
 Such a potential can appear   in the approximation of the
  1-loop finite temperature effective potential of 
the two interacting massive scalar fields.

\begin{eqnarray}
V(\phi_1,\phi_2,T) &=& (D_1T^2-\mu_1^2)\phi_1^2+\lambda_1\phi_1^4
 \nonumber\\
&+&(D_2T^2-\mu_2^2)\phi_2^2+\lambda_2\phi_2^4+g \phi_1^2\phi_2^2 +C,
\label{V}
\end{eqnarray}

where $-\mu_i^2 \phi_i^2 (i=1,2)$ is  the bare 
 mass term of $\phi_i$, and $D_i$  is the  
coefficient of thermal
 mass correction term\cite{dolan}. 
Here  the constant $C$  is introduced to make the
 cosmological constants zero.

 The mutual
 interaction term: 
\beq
V_{int}=g\phi_1^2\phi_2^2.
\eeq
is essential for inverse symmetry breaking.
 We will consider the case where this term exists
 in the tree level potential. This term may also arise 
via fermion exchange box diagrams, even if it is absent
 in the tree level potential \cite{g}.

When the fields have v.e.v.(vacuum expectation value), 
they acquire additional masses through $V_{int}$. 
Their effective masses squared at a temperature $T$
 without $D_iT^2$ terms are 
\begin{eqnarray}
m_{1eff}^2(T)&\equiv &2(-\mu_1^2+g \langle\phi_2(T)\rangle^2),\\
m_{2eff}^2(T)&\equiv &2(-\mu_2^2+g \langle\phi_1(T)\rangle^2),
\label{meff}
\end{eqnarray}
 
where  $\langle\phi_i(T)\rangle$  is the v.e.v. of $\phi_i$  at $T$.

 Then the phase transition temperature 
$T_{ci}$ at which the coefficient of $\phi_i^2$
 vanishes can be defined, $i.e.,$
\beq
T_{ci}^2 \equiv -\frac{m_{ieff}^2(T_{ci})}{2D_i}. 
\eeq

From now on we will consider the case 
\beq
 T_{c1} > T_{c2},
\label{isb1}
\eeq

which means that $\langle\phi_1(T)\rangle$ becomes nonzero at $T_{c1}$,
  and after the expansion of the universe  $\langle\phi_2(T)\rangle$
 becomes nonzero  at the lower temperature $T_{c2}$ in turn.
 If $\langle\phi_2(T)\rangle$ is sufficiently large at $T_{c2}$, 
the symmetry of $\phi_1$ broken at $T_{c1}$ can be
 restored due to the additional mass term 
from $V_{int}$(see eq.(\ref{meff}) and Fig.1.).
This is so-called `Inverse symmetry breaking'\cite{isb}.

If at this temperature($T_{c2}$) $\phi_1$ rolls down slowly 
from $\langle\phi_1(T_{c2})\rangle$ to zero and its energy dominates 
others, we can expect a chaotic type slow-rollover 
inflation and regard  $\phi_1$ as an inflaton field.
Note that here we use the terminology `chaotic' to mean
 a kind of inflation potential and not chaotic initial
 condition\cite{lecture}.

$\langle\phi_i(T)\rangle$ can be found from the relation 
$dV(\phi_1,\phi_2,T)/d\phi_i=0$.
 From eq.(\ref{V}) one can obtain
\beq 
\langle\phi_1(T)\rangle=\sqrt{\frac{\mu_1^2-g\langle\phi_2(T)
\rangle^2-D_1T^2} {2\lambda_1}}\simeq
\sqrt{\frac{\mu_1^2}{2\lambda_1}}\equiv\sigma_1,
\label{sigma1}
\eeq
when $T_{c2}< T<T_{c1}$.

The above approximation is justified by the facts that 
$\langle\phi_2(T)\rangle=0$ in this temperature range,
and $D_1 T^2$ term decreases rapidly
 after the phase transition at $T_{c1}$( see again Fig.1. ).

And similarly when $T<T_{c2}$,
\beq 
\langle\phi_2(T)\rangle \simeq
\sqrt{\frac{\mu_2^2}{2\lambda_2}}\equiv\sigma_2.
\label{sigma2}
\eeq

From now on, to simplify the calculation, 
we will use $\sigma_i$ 
 as an approximation of $\langle\phi_i(T)\rangle$ in
the temperature region described above. 
It is a good enough approximation for the order of magnitude estimates.

%And by the same reason we approximate $V(\phi_1,\phi_2,T)$
%as $V(\phi_1,\phi_2,0)$ when $T\leq T_{c2}$. 

Note that $\sigma_1$ and $\sigma_2$ minimize $V(\phi_1,0,0)$ and 
$V(0,\phi_2,0)$ respectively.

\section*{ III. CONSTRAINTS FOR THE INFLATION }
\vskip 2.0mm

\hspace{.5cm}
In this section the conditions for
a successful inflation will be
obtained.
There are many constraints for the
  successful inflation models.
The most significant one comes from the density perturbation:
\beq
[\frac{\Delta T}{T}]^2_Q=\frac{32\pi V_{inf}^3}{45V_{inf}^{'2}M_P^6},
\eeq
where $V_{inf}'$ is $\frac{dV_{inf}}{d\phi_1}$ at the horizon crossing of the 
observed scale. 
We consider the quadratic term  dominated inflaton 
potential $V_{inf}\equiv m_1^2\phi_1^2/2$ which is the
$\phi_1$ dependent part of  the approximation of 
$V(\phi_1,\phi_2,T \leq T_{c2})$.
So $m_1^2/2 \simeq -\mu_1^2+g\sigma_2^2$.

COBE\cite{cobe} observation, $ [\frac{\Delta T}{T}]_Q
 \simeq 6\times 10^{-6} $, demands $m_1\simeq 10^{13}GeV$
 for our model. 
 
Sufficient expansion condition requires\cite{book}
\beq
\sigma_1=\sqrt{\frac{N}{2 \pi}} M_P \stg 3 M_P
\label{sufficient}
\eeq
for $ e^N$ expansion and $ N\stg 60$.\\
Note that for the quadratic term dominated inflaton potential
slow-rolling condition $m_1 \ll H$ is automatically satisfied 
for $ \sigma_1 \stg  M_P$.
The above two constraints are common to many mass term dominated 
chaotic type inflation models. 

Now we will investigate conditions specific
for our model.
 
First, the condition for inverse symmetry breaking(eq.(\ref{isb1}))
 is equal to 
\beq
\frac{\mu_1^2}{D_1}>\frac{\mu_2^2-g\sigma_1^2}{D_2}.
\label{isb2}
\eeq

Second, the phase transition at $T_{c2}$ must be energetically favorable
 to take place.
It means that the free energy released by symmetry
 breaking by $\phi_2$  
must be larger than the free energy absorbed by 
symmetry restoration by $\phi_1$.
This implies  
\beq
V(\langle\phi_1(T)\rangle,0,T \ge T_{c2})
 - V(0,\langle\phi_2(T)\rangle
,T \le T_{c2}) > 0,
\label{potential2}
\eeq

or approximately
$V(\sigma_1,0,0)
 - V(0,\sigma_2,0)  > 0$, which is equivalent to

\beq
\mu_1^2 \sigma^2_1 < \mu_2^2\sigma_2^2.
\label{potential}
\eeq

Third, restoring the symmetry of $\phi_1$ implies $m_{1eff}^2(0)>0$, or 
\beq
\mu_1^2<g\sigma^2_2.
\label{restor}
\eeq
Similarly, the broken symmetry of $\phi_2$ implies $m_{2eff}^2(0)<0$, 
\beq
\mu_2^2>g\sigma^2_1.
\label{broken}
\eeq

From eq.(\ref{isb2}) and eq.(\ref{potential}) we obtain
\beq
D_1<\frac{\mu_1^2}{\mu_2^2-g\sigma_1^2}D_2
\simeq\frac{\mu_1^2}{\mu_2^2}D_2
<(\frac{\sigma_2}{\sigma_1})^2 D_2,
\label{D1}
\eeq
where we have used eq.(\ref{broken}) in the approximation.

And finally, we want the potential $V_{inf}$ to be dominated by $\phi_1^2$ term 
 rather than by $\phi_1^4$ term. So
\beq
\mu^2_1<\frac{2}{3}g \sigma^2_2.
\label{quad}
\eeq
Using eq.(\ref{sigma1}) and eq.(\ref{sigma2}),
one can rewrite the constraints(eq.(\ref{isb2}) and eq.(\ref{potential}))
 with $\lambda_i$ instead
of $\mu_i$.
\beq
\lambda_1(\frac{\sigma_1}{\sigma_2})^4 <\lambda_2<
\lambda_1(\frac{\sigma_1}{\sigma_2})^2 \frac{D_2}{D_1}+
\frac{g}{2}(\frac{\sigma_1}{\sigma_2})^2. 
\label{lambda2}
\eeq

Let us further consider miscellaneous constraints.
One loop correction to $\lambda_i$ should not be larger than itself,
 i.e., $\lambda_i \stg 0.1 g^2$.

  Whether $\phi_2$ drives an inflation  or not at $T_{c2}$, $\phi_2$ 
oscillates
  around the potential minima($\sigma_2$) 
 with period $\sim 1/m_2$
after the phase transition (see fig.1),
 and its energy density $\rho_{\phi_2}$
  decreases as $R^{-3}(t)$ like classical 
nonrelativistic matter field\cite{osc}.
 Here $m_2^2/2\equiv -\mu_2^2+g\sigma_1^2$ is an approximation
of $m_{2eff}^2(T)$ at $T_{c2}\le T <T_{c1}$.

Since $R\propto t^{2/3}$ in the matter dominated era,
$\rho_{\phi_2}$ is proportional to $t^{-2}$ during the
oscillation.
(Even if $\rho_{\phi_2}$ rapidly changes to
radiation energy so that the universe is in radiation-dominated
era, the energy density is proportional to $t^{-2}$ and 
the above arguments still hold.)

We need to know the time ($\triangle t_{osc}$)
when $\rho_{\phi_2}$ decreases
to $\rho_{\phi_1}$ and the inflation by $\phi_1$ begins.
From the fact that $ \rho_{\phi_2}(t)\simeq
 \rho_{\phi_2}(t_2)H(t_2)^{-2}t^{-2}$ 
this time scale is given by
\beq
\triangle t_{osc}\simeq \frac{1}{H(t_2)}
[\frac{\rho_{\phi_2}(t_2)}{\rho_{\phi_1} }]^{\frac{1}{2}}
\sim \frac{M}{m_1\sigma_1},
\eeq
where $M\equiv M_P/8\pi$ is the reduced Planck mass, $t_2$  is the time when
 the oscillation of $\phi_2$ starts and
$H(t_2)\sim m_2\sigma_2/M\sim \rho_{\phi_2}^{\frac{1}{2}}/M$. We have also used the fact that
  $\rho_{\phi_i}\sim m_i^2 \sigma_i^2$ before $\phi_i$
start to oscillate.

During $\triangle t_{osc}$, 
$\phi_1$ should not fall down too much. 
Since the equation for $\phi_1$ is
\beq
 3H\dot{\phi_1}= -m_1^2\phi_1, 
\eeq
whose solution is $\phi_1=\sigma_1-m_1M_Pt/2\sqrt{3}$\cite{book},
 the rolling time scale  is 
$\triangle t_{rol}\sim 1/m_1$.( the dots denote time derivatives.) 
Therefore one can know that if $\sigma_1 \gg M$,
 $\triangle t_{osc} \ll \triangle t_{rol}$ and
$\phi_1$ does not decrease too
 much during $\phi_2$ oscillation, and one could  expect
 the inflation by $\phi_1$.

\section*{IV. AN EXAMPLE WITH $SU(5)$ GUTs AND NUMERICAL STUDY }

\hspace{.5cm} Let us apply our
model to $SU(5)$ GUTs.
 Consider the case where $\phi_2$ is a 
$SU(5)$ Higgs field\cite{new}.
Then the phase transition temperature
 $T_{c2}\simeq 10^{15}GeV\simeq\sqrt
{(\mu_2^2-g\sigma_1^2)/D_2}\simeq
  \sqrt{\mu_2^2-g\sigma_1^2}$,
because $D_2=\frac{75}{8}g^2_{SU(5)}\simeq 3$
with the unified  gauge coupling $g_{SU(5)}$.

We also know that $\sigma_2\simeq M_X/g_{SU(5)}
\simeq 10^{15}GeV$. 

From eq.(\ref{sufficient}) and eq.(\ref{D1}) 
it is easy to  find that
\beq 
D_1 <(\frac{\sigma_2}{\sigma_1})^2 D_2
\stl 10^{-8}.
\eeq
From the density perturbation constraint $m_1^2/2\simeq
(10^{13}GeV)^2 \simeq -\mu_1^2+g\sigma_2^2\leq g\sigma_2^2$ 
we get $g\stg 10^{-4}$.
However $D_1\simeq 0.1g< 10^{-8}$, so $g< 10^{-7}$. Hence 
 $g$ can not satisfy the both conditions. 
This problem is easily solved by considering the GUTs
 models whose energy scale is larger ($T_{c2}\simeq
 10^{16} GeV$).
 In this case, using the same procedure we obtain $g\stg 
10^{-6}$ and $D_1\stl 10^{-6}$, so all the condition is 
satisfied within our approximation.

From eq.(\ref{restor}) and eq.(\ref{sigma1})  we obtain 
\beq
\lambda_1<\frac{g}{2} (\frac{\sigma_2}{\sigma_1})^2\stl 10^{-12},
\label{lambda1}
\eeq
 so $\lambda_1\ll g$. 

Such a small coupling constant is typical to
many slow-rollover inflation models, and gives rise to
a thermal non-equilibrium problem.
Like many other slow-rollover inflation models
except for the chaotic inflation model, it is very hard
to establish initial thermal equilibrium required
for our model.

For the following, we will assume that somehow 
this equilibrium is established
and $\phi_i$ has the appropriate initial values. 
( The parametric resonance mechanism\cite{parametric}
may help good reheating, but it is still unclear that
produced light particles can obtain the thermal equilibrium
before $T_{c2}$.)

If we want any inflation at $T_{c2}$, the vacuum energy 
of $\phi_1$ or $\phi_2$
must be larger than the radiation energy.
In this case,  from eq.(\ref{potential}) the energy of $\phi_2$ is
 larger than that of $\phi_1$, so it is possible that 
there is a new inflation by `$\phi_2$' before that
 by $\phi_1$. 
So our model could be a kind of `double inflation'\cite{double}.

Whether the first inflation(by $\phi_2$) can exist depends 
on the rolling speed of $\phi_2$ at this phase transition. 
 Since the number of e-foldings of expansion in the
 new inflation is given by $N\simeq(H/m_2)^2$, the first
 slow-rollover inflation is available only for $m_2\ll H$.

However, from the fact that $m_2^2/2=-\mu_2^2+g\sigma_1^2$,
one can know that  $m_2\gg H_2\simeq 10^{13} GeV$ without fine tuning
  and there is no slow-rollover inflation driven by 
$\phi_2$ preceding $\phi_1$ inflation with GUTs.

Now we will discuss the numerical study of our model.
The process of our inflation model seems to be rather complicated.
To confirm the scenario we perform numerical study
of following equations for the evolution of the fields:
\beqa
H&=&[\frac{1}{3M^2}(\frac{\dot{\phi_1}^2}{2}+\frac{\dot{\phi_2}^2}{2}
+V)]^{\frac{1}{2}}, \nonumber \\
\ddot{\phi_i}&+&3H\dot{\phi_i}+\frac{\partial V}{\partial \phi_i}=0,
\eeqa
where $V$ is $V(\phi_1,\phi_2,0)$ in eq.(\ref{V}).
We have ignored thermal contributions which may become small
relatively
when there is inflation or oscillation of $\phi_1,\phi_2$.

Fig.2 shows the results with $m_1=10^{13}GeV,
m_2=5 \times 10^{16}GeV, \sigma_1=5M, \sigma_2=5\times 10^{-2}M$
and $g=10^{-7}$.  

After the long oscillation  of $\phi_2$ for $\tau\stg 11$ (in realistic case,
this oscillation disappears rapidly by producing particles.),
$\rho_{\phi_2}$ decreases and $\phi_1$ rolls down
and begin the inflation. The sign of the inflation by $\phi_1$ can be
identified by the flat region of $H$ graph( $\tau \stg 16$).
After the inflation ends, $\phi_1$ starts to oscillate when $\tau\simeq 19$.
 
Now let us consider the case where no initial thermal
equilibrium state is established.  
It is well known that at the Planck scale the typical initial value of $\phi_1$
could be about $\lambda_1^{-1/4} M_P\gg M_P$.
Hence, generally there could be an chaotic inflation by $\phi_1$
before the inflation  by $\phi_1$ and/or by $\phi_2$
at the lower temperature.

Whether there has been a chaotic
inflation or not, $\phi_1$ field rolls down to $\sigma_1$ and start to oscillate
when $\phi_1-\sigma_1$ becomes about $M_P$.
Since $m_2 \gg m_1$, during the chaotic inflation
$\phi_2$ rolls down to $\sigma_2$ rapidly, then the effective
mass of $\phi_1$ becomes positive and $\phi_1$
may roll  down to zero again.
In this case our scenario is hardly
distinguishable from the ordinary chaotic inflation
by $\phi_1$.
So it seems to be essential to assume the initial thermal equilibrium,
if we consider our model with GUTs.

\section*{V.  DISCUSSIONS}
\hspace{.5cm} 
The most special feature of our model is that we can choose 
the initial value of the inflaton field($\sigma_1$)
 by varying the parameters.

From eq.(\ref{lambda2}) and eq.(\ref{lambda1})
we know that the relation
 $\lambda_1\ll g <\lambda_2$ should be
satisfied for the successful inflation.

 For some parameter ranges
  our model could be a 
 two-filed double inflation whose properties depend
 on the rolling speed of $\phi_2$.

Our model with the GUTs phase transition  requires 
the GUTs energy scale  to  be  $O(10^{16}GeV)$,
while assumption of thermal equilibrium is needed
like many other slow-rollover inflation models.

The numerical study indicates that
in spite of complexity of out model inflation
could occur with parameters constrained by many conditions. 

This model may also be used to give the appropriate 
density perturbation to match
  COBE normalization with galaxy-galaxy correlation
 function\cite{double_density}.
Note that for this purpose 
$\sigma_1$( eq.(\ref{sufficient})) should
be lowered so that we can observe
the effect of the inflation by $\phi_2$.

 Many constraints on the masses and couplings of the
 fields for the successful inflation and inverse
 symmetry breaking are studied.
However, some of the requirements can be abandoned.
  For example, $\phi_1$ needs not have zero v.e.v. 
after inflation and may have some finite v.e.v.
 In this case, $\phi_1$ could be a scalar field
 responsible for the broken symmetry in some 
particle physics theories.

It is also possible that  inflaton potential is
 dominated by quartic term not by quadratic term.

Furthermore, for more  general case  the potential 
$V(\phi_1,\phi_2,T)$  
may have  small barrier term such as $T\phi_i^3$. 
 In this case, it is possible that there is a first order
 inflation by $\phi_2$ which is interesting, because 
it could 
be  another mechanism for the recently proposed open 
inflation models\cite{open}.      

In a word, there still remain  various scenarios  to be 
studied in different parameter spaces in this model
 where the new way of onset of inflation is introduced.

\vskip 1cm
\section*{ ACKNOWLEDGMENTS }
The authors are grateful to H. Kim,  H. Kwon and Y. Han for
 useful comments.
 This work was supported in part by KOSEF.

\newpage
\section*{ Figure Caption }

Fig.1. Schematic diagram for inverse symmetry breaking.\\
$\langle\phi_1(T)\rangle$(thick line) and 
$\langle\phi_2(T)\rangle$(dashed line) versus
temperature $T$.

\vspace{1cm}
Fig.2. The results of numerical study showing the evolution
 of $\phi_1,\phi_2$ and $H$ versus time in log scale
$\tau=ln(m_2 t)$. 
$\phi_1$ is in units of $M$, $\phi_2$ in units of $10^{-2}M$
and $H$ in units of $10^{-2} m_2$.


\vskip 5.4mm
\newpage





\begin{thebibliography}{99}
\bibitem{guth} A. Guth, {\it Phys. Rev.,} {\bf D23}, 
347(1981); \\
 For a review see: K. Olive, {\it Phys. Rep.,} {\bf 190},
 307 (1990) 

\bibitem{liddle}  A. Liddle and D. Lyth, {\it Phys. Rep.,} 
{\bf 231}, 1 (1983) 

\bibitem{new}  A. Linde, {\it Phys. Lett.,} {\bf B108},
 389 (1982); 
 A. Albrecht and P. Steinhardt, {\it Phys. Rev. Lett.,}
 {\bf 48}, 1220 (1982), 
 P. Steinhardt and M. Turner, {\it Phys. Rev.,} {\bf D29},
 2162 (1984) 


\bibitem{su5}  Q.Shafi and A.Vilenkin, {\it Phys. Rev. Lett.,} 
{\bf 52}, 691 (1984) 

\bibitem{chaotic} 
 A. Linde, \PLB {\bf 129}, 177 (1983); 

\bibitem{isb} R. Mohapatra, {\it Phys. Rev. Lett.,} {\bf 42}, 1651 (1979),
{\it Phys. Rev.,} {\bf D20}, 3390 (1979);
   S. Weinberg, {\it ibid.,} {\bf D9}, 3357 (1985), 
 T. Farris and T. Kephart, {\it J. Math. phys.,} {\bf 32},
 2219 (1991) 


\bibitem{pi}  P. Langacker and S.
 Pi,{\it Phys. Rev. Lett.,}
 {\bf 45}, 1 (1980) 

\bibitem{anti}  A. Masiero and D. Nanopoulos,
{\it Phys. Lett.,}
 {\bf B138}, 91 (1983); 
 A.love and S.J.Stow, {\it Nucl. Phys.,} 
{\bf B257},271 (1985) 

\bibitem{two}  J. Silk and M. Turner, {\it Phys. Rev.,}
 {\bf D35}, 419 (1987); 
 L. Kofman and A. Linde, {\it Nucl. Phys.,} {\bf B282},
 555 (1987), 
 F. Adams and K. Freese, {\it Phys. Rev.,} {\bf D43},
 353 (1990) 

\bibitem{hybrid} A. Linde, {\it Phys. Lett.,}
 {\bf B249}, 18 (1990), 
 {\it ibid.,} {\bf B259}, 38 (1991); 
 E. Copeland, A. Liddle, D. Lyth, etal.,{\it Phys. Rev.,} {\bf D49},
 6410 (1994), 
 Y. Wang .,{\it ibid.,} {\bf D50}, 6135 (1994); 
 E. Stewart. ,{\it Phys. Lett.,} {\bf B345}, 414 (1995) 

\bibitem{dolan}  L. Dolan and R. Jackiw, {\it Phys. Rev.,}
 {\bf D9}, 3320 (1974) 

\bibitem{g} A. Salam and V. Elias, {\it Phys. Rev.,} 
{\bf D22}, 1469 (1980) 

\bibitem{lecture}  A. Linde,
{\it Lectures on Inflationary Cosmology, } 
preprint SU-ITP-94-36(Oct 1994), hep-th/9410082

\bibitem{cobe}  G. Smoot etal., {\it Astrophys. J.,}
 {\bf 396},L1 (1992) 

\bibitem{book}  A. Linde,
{\it Particle physics and inflationary cosmology},
 (harwood, New-Work,1990) 


\bibitem{osc}  M. Turner, {\it Phys. Rev.,}
 {\bf D28}, 1243 (1983) 

\bibitem{parametric}  L. Kofman, A. Linde, and
A. Starobinsky, {\it Phys. Rev. Lett.,}
 {\bf 73}, 3195 (1994) 

\bibitem{double} J. Silk and M. Turner, {\it Phys. Rev.,}
 {\bf D35}, 419 (1987), 
 R. Holman, E. Kolb, S. Vadas, and Y. Wang.,
 {\it Phys. Lett.,} {\bf B269}, 252 (1991); 
 L. Kofman, A. Linde and  A. Starobinsky,{\it ibid.,}
 {\bf B157}, 361 (1985) 

\bibitem{double_density}  L. Kofman and D. Pogosyan.,
{\it Phys. Lett.,} {\bf B214}, 508 (1988); 
 P. Peter, D. Polarski and A.
 Starobinsky., {\it Phys. Rev.,}
 {\bf D50}, 4827 (1994) 

\bibitem{open}  J. Gott., {\it Nature.,} {\bf 295},
 304 (1982); 
 A. Linde., {\it Phys. Lett.,} {\bf B351}, 99, (1995);
 M. Bucher., A. Goldhaber and N. Turok., preprint 
PUPT-1527 ( Jan 1995),  hep-ph/9501396,
\end{thebibliography}
\end{document}